\documentclass{article}
\usepackage{spconf,amsmath,graphicx}
\PassOptionsToPackage{hyphens}{url}
\usepackage{hyperref}
\usepackage{xcolor}
\hypersetup{colorlinks=true}
\usepackage{cite}
\usepackage{ifthen}

\title{Synth2Aug: cross-domain speaker recognition \\
with TTS synthesized speech}
\ifthenelse{\isundefined{\makeblind}}
{
\name{Yiling Huang$^{*}$, Yutian Chen$^{\dagger}$, Jason Pelecanos$^{*}$, Quan Wang$^{*}$}
\address{$^{*}$Google LLC, USA \qquad $^{\dagger}$DeepMind, London, UK}
}
{
\name{BLIND}
\address{BLIND}
}

\begin{document}
\ninept
\maketitle
\begin{abstract}
In recent years, Text-To-Speech (TTS) has been used as a data augmentation technique for speech recognition to help complement inadequacies in the training data. Correspondingly, we investigate the use of a multi-speaker TTS system to synthesize speech in support of speaker recognition. In this study we focus the analysis on tasks where a relatively small number of speakers is available for training. We observe on our datasets that TTS synthesized speech improves cross-domain speaker recognition performance and can be combined effectively with multi-style training. Additionally, we explore the effectiveness of different types of text transcripts used for TTS synthesis. Results suggest that matching the textual content of the target domain is a good practice, and if that is not feasible, a transcript with a sufficiently large vocabulary is recommended.


\end{abstract}

\section{Introduction}
\label{sec:intro}

Speaker recognition is the process of recognizing the identity of the speaker from a spoken utterance. Recent work has explored different deep neural network (DNN) architectures~\cite{snyder2017deep, snyder2018x, variani2014deep, chen2015locally, li2017deep, zhang2016end, heigold2016end, wan2018generalized} to produce compact speaker embeddings. In this paper, our work is based on the d-vector model described in~\cite{wan2018generalized} for the text-independent speaker recognition task.

DNN speaker recognition has benefited from the use of various data augmentation techniques~\cite{snyder2018x, yamamoto2019speaker, huang2019exploring, mohammadamini2020data}. While speech recognition has also used such techniques~\cite{lippmann1987multi,ko2017study,kim2017generation}, some recent work has explored the use of TTS for supplementing speech recognition training data~\cite{laptev2020you, gokay2019improving, wang2020improving}. These advances are enabled by the fact that synthesizing close-to-human-quality multi-speaker speech has become more mature~\cite{oord2016wavenet, shen2018natural, ping2018deep, jia2018transfer, chen2019sample}.

Likewise, the speaker recognition community has started to explore the use of TTS, but mainly on anti-spoofing. Researchers have worked on determining if speaker recognition systems were vulnerable to TTS techniques and, if so, examining ways to make such systems robust to TTS spoofing attacks~\cite{simonchik2014stc,shchemelinin2013examining, shchemelinin2014vulnerability,kinnunen2017asvspoof,wang2020asvspoof}. While the speech recognition community found some utility in using TTS for data augmentation, we are currently not aware of studies examining TTS for speaker recognition data augmentation. This paper begins to address this gap.

From a speaker recognition perspective, a speech recording can be abstracted into two sources of variability; speaker variability and channel/environment effects. Many data augmentation techniques seek to expand upon the variability observed for channel/environment effects. There is limited work regarding how speaker variability can be augmented. One example is that the time warping component of the SpecAugment technique~\cite{park2019specaugment, faisal2019specaugment, wang2020specaugment} could be considered as a basic form of speech rate perturbation. Interestingly, multi-speaker TTS synthesis seems to be a natural candidate for providing additional speaker-centric variability.

We narrow the scope of this initial work to make performance analysis more tractable (while beginning with perhaps an easier task). In this paper, we study how speech synthesis can be leveraged to improve cross-domain speaker recognition performance with the proposed Synth2Aug approach. As part of this work we assess if TTS as a data augmentation method can provide additional information over a carefully tuned multi-style training setup. We also explore if speech content plays a role in the effectiveness of TTS synthesis for speaker recognition. Note that our experiments are conducted under two main constraints: there is no target domain training data, and our training data consist of a relatively small group of speakers.


The rest of the paper is organized as follows. Section~\ref{sec:system_description} gives a brief introduction on our Synth2Aug approach, followed by detailed explanations of the speaker recognition and TTS systems. In Section~\ref{sec:experiments}, we present two sets of experiments; one studying the contribution of TTS synthesis to speaker recognition performance and the other exploring the effect of TTS textual content. This is followed by conclusions and future work in Section~\ref{sec:conclusions}.

\section{Description of Proposed System}
\label{sec:system_description}

\begin{figure*}
\centering
\includegraphics[width=0.8\textwidth]{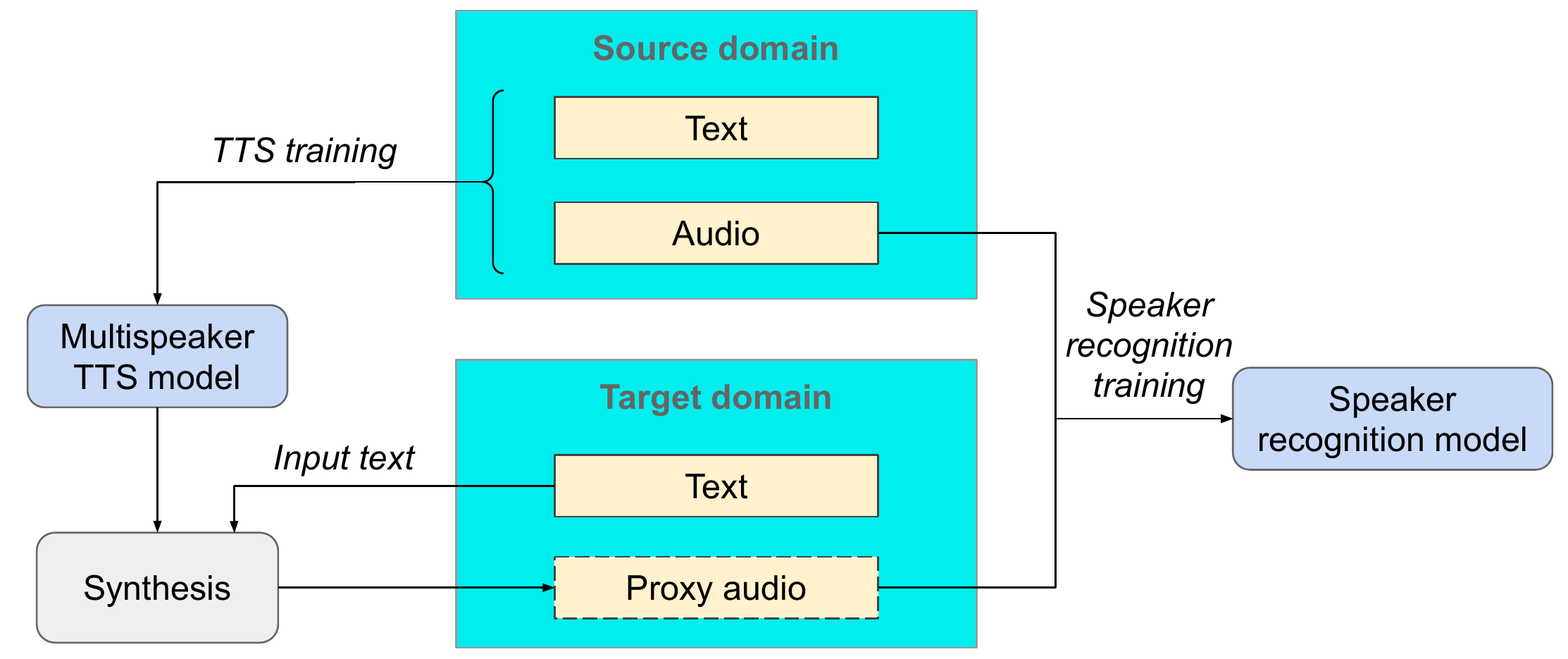}
\caption{Diagram of the proposed Synth2Aug system which uses TTS synthesized speech as augmented data. In an ideal situation the synthesis system would generate speech (\textit{proxy audio}) fully representative of the target domain in all aspects. However, in practice, while it may be possible to learn and represent the textual information of the target domain, the acoustic properties of the signal (e.g. channel effects) would be difficult to capture.}
\label{fig:workflow}
\end{figure*}

As mentioned previously, the goal is to build a speaker recognition system that leverages TTS synthesis to potentially improve performance on cross-domain data. 
While it may be difficult to acquire multi-speaker audio training data for the target domain, it is relatively simple to use similar textual content for speech synthesis such that it better relates to the target domain. We propose Synth2Aug to overcome the cross-domain challenge, as illustrated in Figure~\ref{fig:workflow}. Given existing source domain training data, we train a multi-speaker TTS model to synthesize new speech data based on the target domain textual content. We incorporate the synthesized speech, along with the existing training data, into speaker recognition model training. In the following sections, we will walk through the details including: the speaker recognition system, the multi-speaker TTS system, and the specifics of the synthesized speech.

\subsection{Speaker recognition system}
Our speaker recognition model is a text-independent, d-vector system similar to the model architecture in~\cite{wan2018generalized}. While the work could easily be applied to both speaker identification and verification, we are performing a speaker verification task for the discussion in the paper. To aid in reproducibility, we share the system details which cover feature extraction, model architecture, system training and evaluation.

The feature extraction process is implemented based on~\cite{prabhavalkar2015automatic}. The audio is first transformed into frames of 25ms width and 10ms shift, followed by the extraction of 40-dimensional log-Mel-filterbank energy features for each frame. A model-based Voice Activity Detection (VAD) component is integrated to exclude non-speech frames. Finally, we stack adjacent frames of features by concatenating two frames of size 40 into a single frame of size 80, which results in a frame sequence of approximately half the number of original frames.

The model is composed of 3 LSTM layers with projections~\cite{sak2014long}. For each LSTM layer, the number of hidden nodes is 768 and the output is projected down to 256 dimensions followed by a tanh activation function. A final linear layer of 256 dimensions is appended to the last LSTM layer. 



For system training, each mini-batch is constructed with 16 randomly selected speakers with 8 randomly selected utterances per speaker. When there are multiple training datasets, we use the MultiReader technique proposed in~\cite{wan2018generalized}. We assign a weight to the batch of utterances fetched from each data source and compute the combined loss. The loss function is the generalized end-to-end softmax loss (implementation is based on~\cite{wan2018generalized}).

For system evaluation, we use the L2-normalized output of the last frame from the last 256-dimensional layer to represent the speaker embedding. This speaker embedding is known as the d-vector. The enrollment embedding is calculated by averaging multiple enrollment utterance d-vectors for each speaker and applying L2-normalization again. The test utterance is compared with the enrollment embedding using the cosine similarity metric. 

\subsection{Text-to-speech model}
\label{sec:tts}

The Text-To-Speech (TTS) model is based on a Tacotron 2 model~\cite{shen2018natural}, which consists of a spectrogram prediction network to convert phoneme sequences to Mel-spectrogram, and a WaveRNN~\cite{kalchbrenner2018efficient} model as a vocoder to convert the Mel-spectrogram to speech waveform. We train a multi-speaker Tacotron 2 model to support multi-speaker speech synthesis. Specifically, every speaker is represented with a $d$-dimensional embedding vector. We feed it into the decoder of the Tacotron model and concatenate it with the output vector from the attention model. These embedding vectors are trained as parameters together with the main model. To clarify, the speaker embedding vector introduced here is a trainable vector bound to the Tacotron model. It is a different concept from the d-vector speaker embedding in speaker recognition. We will always refer to the speaker embedding in speaker recognition directly as ``d-vector" in this paper.

In addition, we introduce a sequence of 32-dimensional latent variables in the decoder (as per~\cite{hsu2018hierarchical}) to capture the residual intraspeaker style variations.
The latent variables at each output step are sampled independently from a mixture of Gaussian priors. We follow the variational autoencoder method~\cite{kingma2014auto}, and train a recognition model from the output Mel-spectrogram to infer the posterior distribution of the latent variables during training.

The embedding space when trained with a large number of speakers is expected to capture the unique characteristics of individual speakers. To further increase the diversity of synthesized voices, we train multiple models with different dimensional speaker embedding spaces (of dimension 32, 64, 128) and synthesize speech from all of them.

The training data for the TTS model are audiobook read speech. It is a combination of two proprietary speech corpora and the LibriTTS dataset~\cite{zen2019libritts}. There are in total 2,218 speakers, including 55 and 1,080 speakers from the proprietary datasets, and 1,083 LibriTTS speakers (clean-100 and clean-360), with an average of around 500 utterances per speaker. The training process is the same as~\cite{shen2018natural}, which involves first training the feature prediction network on its own, followed by the WaveRNN training independently on the outputs generated by the feature prediction network.

\subsection{Synthesized speech}
\label{sec:synthesized-speech}

When generating the synthesized speech, we consider two approaches for creating a diverse set of voices.

The first approach is to directly use the embedding vectors of the 2,218 training speakers from the TTS model with the 128-dimensional embedding space. The model synthesizes utterances according to each corresponding embedding. Note that we do not include the other two models (i.e. 32d and 64d) because the underlying speakers are the same. This approach preserves the genuine characteristics of the speakers with high quality. However, the speaker variability is relatively limited due to us reusing the existing real speakers.


The second approach is to synthesize artificial voices by sampling the speaker embedding in the learned embedding space. An arbitrarily sampled vector in the speaker embedding space may be far from any realistic training voices and this leads to a generalization challenge for the TTS model. To mitigate this, we first fit a mixture of 3 full-covariance Gaussians to the set of training speaker embedding vectors, and then draw samples from the Gaussian Mixture Model. This allows us to extract as many voices as needed with a slightly degraded quality, compared to using the speaker embeddings from real training speakers. As the purpose of data augmentation is to improve the robustness of speaker recognition, the additional noise in the utterance is less of a concern.

However, in the second approach, if two embedding vectors are close to each other, the corresponding synthesized voices will be similar as well. To verify this, we first sample a large set of artificial voices, $\mathcal{S}$, from all three TTS models of different dimensions, then compute d-vector cosine similarities with a pre-trained speaker recognition model $\mathcal{M}$\footnote{$\mathcal{M}$ is a standalone speaker recognition model trained separately, solely for the purpose of selecting distinct voices across different dimensional TTS models. This model $\mathcal{M}$ has not seen any training data used in the following experiments.}. Specifically, for each sampled voice we synthesize $N=100$ utterances, obtain the d-vector of each utterance from running the model $\mathcal{M}$, and then compute the average. We find that while the average cosine similarity between real training voices and sampled voices is only about 0.185, the average similarity between two sampled voices is almost 0.5. The intraspeaker variation in such a case could dominate the interspeaker variation. 
In order to avoid this problem, we conduct a ``speaker selection" process, which aims to build a speaker set where the voices are reasonably distant from each other. We incrementally build a subset of speakers $\mathcal{T} \subseteq \mathcal{S}$. At each step, a speaker $i$ is randomly drawn from $\mathcal{S} \backslash \mathcal{T}$, and added to $\mathcal{T}$ if the cosine similarity between $i$ and each $j$ of the already selected speakers does not exceed 0.4, as illustrated in Equation~\ref{eq:cos}.


\begin{equation}
    \cos\left(\frac{1}{N}\sum_{n=1}^N \mathbf{d}_{in}, \frac{1}{N}\sum_{n=1}^N \mathbf{d}_{jn}\right) \leq 0.4,\quad \forall j \in \mathcal{T},
    \label{eq:cos}
\end{equation}

where $\mathbf{d}_{in}$ is the d-vector computed from the $n$th synthesized utterance of the $i$th voice.



We construct this speaker set $\mathcal{T}$ in a greedy way, by searching through sampled voices from the TTS models of embedding dimension $d=32$, then $64$ and $128$ subsequently. As a result, thousands of distinct speaker embeddings are sampled from each model, as shown in Table~\ref{tab:num_speaker}.

\begin{table}
\caption{Number of speakers associated with each synthesized dataset.}
\label{tab:num_speaker}
\begin{center}
  \begin{tabular}{ l | c }
    \textbf{Dataset} & \textbf{Number of Speakers} \\ \hline
    128d-Real & 2,218 \\
    \,\,\,32d-Sampled & 3,180 \\
    \,\,\,64d-Sampled & 2,105 \\
    128d-Sampled & 1,196 \\
  \end{tabular}
\end{center}
\end{table}

For each speaker, real and sampled, we synthesize 10,000 speech utterances. The underlying transcript of the TTS synthesis could contain any text. In Synth2Aug, we collect transcript text which closely matches the target domain text. (We also experiment with different types of transcripts in Section~\ref{sec:mismatch}.)


\section{Experiments}
\label{sec:experiments}

\subsection{Overview}

Before discussing the experiments, we identify some of the design choices made with regard to data selection and the simplification of experiments. When deciding on the training data for both TTS and speaker recognition systems, we constrain ourselves to using the TTS training data for both tasks to avoid unintended side effects and complexity introduced by multiple training datasets. It is easier for the speaker recognition system to use the TTS system training data than the other way around, because the TTS system training requires high quality speech.

We configure two sets of experiments to better understand how speech synthesis can support speaker recognition.

In the first set of experiments, we explore if TTS synthesis could be valuable for our cross-domain speaker recognition problem. In particular, it is important to understand if it can improve upon a well-tuned multi-style training (MTR) setup~\cite{lippmann1987multi,ko2017study,kim2017generation}. The MTR augmentation introduces room reverberation with noise sources including cafes, cars, and ambient noise from quiet environments, as well as music or other types of sounds. The signal-to-noise ratio (SNR) is drawn from a uniform distribution between 3dB and 15dB.

In the second set of experiments, we endeavor to understand how the textual content of the synthesized speech affects the performance of a speaker recognition system trained on this data. We prepare different types of transcripts, including numeric digit sequences, random word concatenation, closely-matched text and exactly-matched text.

\subsection{Evaluation data}
\label{sec:eval-data}
The evaluation data for all the experiments are the vendor collected speech query data. This dataset was collected via a web service, where 70\% of the queries were recorded with mobile phones and 30\% with laptops in relatively quieter environments. Gender is balanced and the ages range from 18 to 45 years old. There are various channel effects because of different types of microphones/devices. The mean test utterance duration is 4.4s, and the standard deviation is 1.6s. The average SNR of this entire test dataset is 18.4dB, calculated by following the WADA-SNR algorithm~\cite{kim2008robust}. The audio prompts are from general speech queries and websites such as wikipedia. Note that these speech queries are completely different from the read speech used for training. As a result, this becomes a cross-domain challenge.

The evaluation utterances are divided into into enrollment and test utterance lists. There are 8,069 utterances from 1,434 speakers in the enrollment list and 194,890 utterances from 1,241 speakers in the test list. For this dataset, we have 192,943 target trials and 200,000 non-target trials.

\subsection{Impact of TTS synthesized speech}
\label{sec:match}

In this section, we prepare various experimental setups to examine the impact of TTS synthesized speech on speaker recognition. The transcript used for speech synthesis is from the 10,000 most popular speech queries. Since our target domain utterances are mainly speech queries, this transcript could be considered as a close representation. We also investigate whether the synthesized speech is effective when used with MTR. The following models are assessed.


\textbf{Baseline}: The baseline training only makes use of the TTS model training data to train the speaker recognition model. There is no data augmentation involved. All setups listed here have these data included for training. We refer to this set of training data as the \textit{Baseline} training data below.

\textbf{TTS-Real}: This corresponds to the approach where, aside from the \textit{Baseline} training data, we additionally include the synthesized speech based on real speaker embeddings using the 128d TTS model. The speaker diversity of the synthesized data is limited because we have a fixed number of real speakers (i.e. 2,218 speakers). We do not apply MTR here.

\textbf{TTS-Sampled}: The alternative to generating speech using real speaker embeddings is to use the sampled embeddings. Here we include all the utterances synthesized from the 32d, 64d and 128d models. New speakers are introduced, and the total number of speakers after speaker selection is much larger than the number of real speakers we had originally. As a result, \textit{TTS-Sampled} also has more utterances than \textit{TTS-Real}. We do not apply MTR here.

\textbf{TTS-Sampled-Small}: In this experiment, we only use the \textit{64d-Sampled} dataset from the 64d TTS model. It has a similar number of speakers to \textit{TTS-Real}. Therefore, the effect of using a different number of speakers is minimized. The main difference is whether we are introducing new speakers, or simply supplementing the existing training data with more utterances for each existing speaker. No MTR is applied.

\textbf{Baseline-MTR}: We augment the \textit{Baseline} training data with MTR to improve coverage. Since this represents an important baseline, it is imperative to optimize its performance. We found that to maximize the performance of the MTR setup we needed to apply MTR to each utterance 15 times. This expands the amount of training data by 15 times as well.

\textbf{TTS-Sampled-MTR}: In this setup MTR is applied to the synthesized speech in \textit{TTS-Sampled}. Both the clean and the MTR synthesized speech are used, along with the \textit{Baseline} training data, to train the speaker recognition model. Note that the \textit{Baseline} training data does not have MTR applied.

\textbf{Combined-MTR}: We combine both MTR and TTS synthesis data augmentation techniques by reusing the optimized data setup from \textit{Baseline-MTR} and appending to it the clean and MTR copies of the synthesized speech data from \textit{TTS-Sampled-MTR}. We have not fine-tuned the weight of each training data source to be optimal in the MultiReader configuration. Instead we assign weights proportional to the size of each training dataset.

\begin{table}
    \caption{Performance of different data augmentation techniques evaluated on vendor provided speech query data. The first column lists models trained with different types of data augmentation techniques. The second column indicates if MTR data augmentation makes up part of the training data. The third column shows the performance of each model.}
    \label{tab:exp1}
    \begin{center}
        \begin{tabular}{ l | c | c }
        \textbf{Model} & \textbf{With MTR?} & \textbf{EER (\%)} \\ \hline
        Baseline  & N & 6.3 \\
        TTS-Real  & N & 5.6 \\
        TTS-Sampled-Small  & N & 4.8 \\
        TTS-Sampled  & N & 4.6 \\ \hline
        Baseline-MTR  & Y & 4.3 \\
        TTS-Sampled-MTR  & Y & 4.0 \\
        Combined-MTR  & Y & \textbf{3.5} \\
        \end{tabular}
    \end{center}
\end{table}

The experimental results are displayed in Table~\ref{tab:exp1}. As shown, using any sort of synthesized data (\textit{TTS-Real}, \textit{TTS-Sampled-Small} or \textit{TTS-Sampled}) in addition to the original training data provides improvement over the \textit{Baseline}. In addition, \textit{TTS-Sampled-Small} outperforms \textit{TTS-Real}. It indicates that extending the existing speaker set with newly sampled voices is helpful. Also, the ability to synthesize new speakers outweighs the slight degradation in voice quality from not using existing real speaker embeddings. Furthermore, when we compare \textit{TTS-Sampled} against \textit{TTS-Sampled-Small}, there is noticeable benefit from introducing additional new speakers.

By analyzing the MTR related experiments, we observe that the optimized \textit{Baseline-MTR} system already shows significant improvement over the non-MTR \textit{Baseline} system. When MTR is applied to the \textit{TTS-Sampled} dataset (\textit{TTS-Sampled-MTR}), the EER is reduced. More interestingly, when these two data augmentations are combined (\textit{Combined-MTR}), the performance is improved further. This shows that TTS synthesis can complement a well-tuned MTR setup. It may be that while MTR addresses the sparsity of channel/environment effects in the speech data, TTS synthesis helps expand speaker-centric variability.



\subsection{Impact of textual content on performance}
\label{sec:mismatch}

The main focus of this section is to assess the impact of different textual content on speaker recognition performance. We synthesize speech with the same number of speakers and utterances, but with different types of transcripts. The transcript contents studied include random digit/word sequences, top speech queries, target domain test dataset text and its word-shuffled variant:


\textbf{Random-Digits}: We prepare text with a very limited vocabulary of numeric digits (zero through nine, plus an additional "oh"). The digits are randomly drawn and concatenated into utterances of 3 to 7 words.

\textbf{Random-Words-100}: The transcript is constructed from the concatenation of random words drawn from a subset of the TIMIT dataset dictionary~\cite{garofolo_1992_1}. The subset consists of 100 words chosen randomly from the full dictionary. Utterances of 3-7 words are created from this subset.

\textbf{Random-Words-Full}: This transcript extends the vocabulary subset in \textit{Random-Words-100} to use all the words in the TIMIT dictionary. We do the same random word concatenation to construct utterances of 3-7 words.

\textbf{Close-Match}: We collect 10,000 of the most popular speech queries as the transcript. The same transcript is used in the previous experiments in Table~\ref{tab:exp1}. We denote this as \textit{Close-Match}, because the texts are close representations of the target domain speech queries.

\textbf{Exact-Match}: In this setup, we directly extract the text from the speech query test dataset. This represents an exact match of the textual content from the target domain. Since the test dataset contains more utterances than what we have in the Section~\ref{sec:match} experiments, we limit the set to 10,000 utterances.

\textbf{Exact-Match-Shuffled}: This transcript is the same as \textit{Exact-Match}, except that there is the additional step of shuffling to reorder the words within each utterance.

\begin{table}
\setlength{\tabcolsep}{4.5pt}
    \caption{Table showing evaluation results and vocabulary sizes for different transcript types. Column \textit{Orig \& TTS} lists the results of the models trained with the original TTS training data included, while column \textit{TTS-only} relates to the models trained with only the synthesized speech. Note that the \textit{Orig \& TTS} column result of the \textit{Close-Match} model is the same as the \textit{Combined-MTR} in Table~\ref{tab:exp1}.}
    \label{tab:exp2}
    \begin{center}
        \begin{tabular}{ l | c | c | c }
          & \textbf{Vocab} & \textbf{Orig \& TTS} & \textbf{TTS-only} \\ 
        \textbf{Model} & \textbf{Size} & \textbf{EER (\%)} & \textbf{EER (\%)} \\ \hline
        Random-Digits  & 11 & 3.7 & 6.9 \\
        Random-Words-100  & 100 & 3.6 & 5.6 \\
        Random-Words-Full & 6,222 & 3.5 & 5.1 \\
        Close-Match  & 3,194 & 3.5 & 4.8 \\
        Exact-Match  & 10,101 & \textbf{3.4} & \textbf{4.3} \\
        Exact-Match-Shuffled  & 10,101 & 3.5 & 4.7 \\
        \end{tabular}
    \end{center}
\end{table}

Now that the individual models are described, we discuss the results in Table~\ref{tab:exp2}. We note that the experiments included here are based on the same configuration as \textit{Combined-MTR} in Table~\ref{tab:exp1}, except that the transcripts being used for TTS synthesis are specific to the models. In the \textit{Orig \& TTS} column of Table~\ref{tab:exp2}, the \textit{Exact-Match} model has the best performance. Although \textit{Random-Digits} performs the worst, it is still relatively comparable to the others. In fact, all the results are within a narrow range. Since the original TTS training data (with MTR applied) already produces a relatively low EER, the expected effect of different textual content is thereby reduced.

In order to fully test the validity of these transcripts, we exclude the original TTS training data, and train our model only on the synthesized data. Results of these revised experiments are shown in the \textit{TTS-only} column of Table~\ref{tab:exp2}.

Not surprisingly, the \textit{Exact-Match} model again has the lowest EER. Here, the synthesized utterances for training share the exact same textual content with the evaluation utterances. The \textit{Close-Match} model outperforms the \textit{Random-Words-Full} model, mostly likely due to the fact that the synthesized data for \textit{Close-Match} is more similar to the evaluation textual content than solely random word concatenations. More interestingly, the \textit{Exact-Match-Shuffled} model, which was trained with semantically meaningless synthesized utterances, is worse than the \textit{Exact-Match} model, but better than the \textit{Close-Match} model by a very small margin. This may indicate that the benefits of matching the target domain text stem from not only similar vocabularies and word frequencies, but also similar word sequences. 

In addition, from \textit{Random-Words-Full} to \textit{Random-Words-100} to \textit{Random-Digits}, there is a clear trend of an increased error rate when the vocabulary size is reduced. Initial observations suggest that a sufficiently large vocabulary (of the order of thousands of unique words) is required to provide close to optimal performance.



In summary of our experiments, it is recommended that speech synthesis uses textual content which: 1) matches the target domain text, and 2) if that is not feasible, uses a comprehensive vocabulary. That said, further work would be required to understand if the same conclusion generalizes to other data sources, data augmentation approaches, or system configurations.




\section{Conclusions and Future Work}
\label{sec:conclusions}

This paper proposed and studied a novel approach, Synth2Aug, to use TTS synthesis as a data augmentation technique for improving cross-domain speaker recognition. When the training data were constrained to a limited number of speakers, using synthesized speech generated from a multi-speaker TTS system improved our cross-domain speaker recognition significantly. TTS synthesis may reduce the sparsity of the training speakers by generating additional artificial speakers. We also found that TTS based augmentation could further improve an optimized MTR augmentation setup.

Additionally, we explored the effect of speech synthesis textual content on model performance. Results indicated that it was helpful to generate transcripts that were matched with the target domain text. Having a sufficiently large vocabulary was also key in the absence of matched transcripts. We also observed that the influence of different textual content was reduced when other non-target domain data was also used in training.





As for future work, one could expand the scope of this work and explore the effect of TTS synthesized data for the in-domain scenario. In particular, it would be useful to understand the performance trade-offs as the quantity of in-domain data increases. From a broader perspective, this paper represents an initial study into how we can use data augmentation to better model the speaker-centric aspects of the speech data. In contrast, past speaker recognition research focused more on environment-centric factors such as channel and noise effects. We challenge the community to examine other ways in which speaker-centric data augmentation can be applied using TTS synthesis or other approaches.



\clearpage

\bibliographystyle{IEEEbib}
\bibliography{refs}

\begin{thebibliography}{10}

\bibitem{snyder2017deep}
David Snyder, Daniel Garcia-Romero, Daniel Povey, and Sanjeev Khudanpur,
\newblock ``Deep neural network embeddings for text-independent speaker
  verification.,''
\newblock in {\em Interspeech}, 2017, pp. 999--1003.

\bibitem{snyder2018x}
David Snyder, Daniel Garcia-Romero, Gregory Sell, Daniel Povey, and Sanjeev
  Khudanpur,
\newblock ``X-vectors: Robust {DNN} embeddings for speaker recognition,''
\newblock in {\em 2018 IEEE International Conference on Acoustics, Speech and
  Signal Processing (ICASSP)}. IEEE, 2018, pp. 5329--5333.

\bibitem{variani2014deep}
Ehsan Variani, Xin Lei, Erik McDermott, Ignacio~Lopez Moreno, and Javier
  Gonzalez-Dominguez,
\newblock ``Deep neural networks for small footprint text-dependent speaker
  verification,''
\newblock in {\em 2014 IEEE International Conference on Acoustics, Speech and
  Signal Processing (ICASSP)}. IEEE, 2014, pp. 4052--4056.

\bibitem{chen2015locally}
Yu-hsin Chen, Ignacio Lopez-Moreno, Tara~N Sainath, Mirk{\'o} Visontai, Raziel
  Alvarez, and Carolina Parada,
\newblock ``Locally-connected and convolutional neural networks for small
  footprint speaker recognition,''
\newblock in {\em Sixteenth Annual Conference of the International Speech
  Communication Association}, 2015.

\bibitem{li2017deep}
Chao Li, Xiaokong Ma, Bing Jiang, Xiangang Li, Xuewei Zhang, Xiao Liu, Ying
  Cao, Ajay Kannan, and Zhenyao Zhu,
\newblock ``Deep speaker: an end-to-end neural speaker embedding system,''
\newblock {\em arXiv preprint arXiv:1705.02304}, vol. 650, 2017.

\bibitem{zhang2016end}
Shi-Xiong Zhang, Zhuo Chen, Yong Zhao, Jinyu Li, and Yifan Gong,
\newblock ``End-to-end attention based text-dependent speaker verification,''
\newblock in {\em 2016 IEEE Spoken Language Technology Workshop (SLT)}. IEEE,
  2016, pp. 171--178.

\bibitem{heigold2016end}
Georg Heigold, Ignacio Moreno, Samy Bengio, and Noam Shazeer,
\newblock ``End-to-end text-dependent speaker verification,''
\newblock in {\em 2016 IEEE International Conference on Acoustics, Speech and
  Signal Processing (ICASSP)}. IEEE, 2016, pp. 5115--5119.

\bibitem{wan2018generalized}
Li~Wan, Quan Wang, Alan Papir, and Ignacio~Lopez Moreno,
\newblock ``Generalized end-to-end loss for speaker verification,''
\newblock in {\em 2018 IEEE International Conference on Acoustics, Speech and
  Signal Processing (ICASSP)}. IEEE, 2018, pp. 4879--4883.

\bibitem{yamamoto2019speaker}
Hitoshi Yamamoto, Kong~Aik Lee, Koji Okabe, and Takafumi Koshinaka,
\newblock ``Speaker augmentation and bandwidth extension for deep speaker
  embedding,''
\newblock in {\em INTERSPEECH}, 2019, pp. 406--410.

\bibitem{huang2019exploring}
Chien-Lin Huang,
\newblock ``Exploring effective data augmentation with {TDNN-LSTM} neural
  network embedding for speaker recognition,''
\newblock in {\em 2019 IEEE Automatic Speech Recognition and Understanding
  Workshop (ASRU)}. IEEE, 2019, pp. 291--295.

\bibitem{mohammadamini2020data}
Mohammad Mohammadamini and Driss Matrouf,
\newblock ``Data augmentation versus noise compensation for x-vector speaker
  recognition systems in noisy environments,''
\newblock {\em arXiv preprint arXiv:2006.15903}, 2020.

\bibitem{lippmann1987multi}
Richard Lippmann, Edward Martin, and D~Paul,
\newblock ``Multi-style training for robust isolated-word speech recognition,''
\newblock in {\em ICASSP'87. IEEE International Conference on Acoustics,
  Speech, and Signal Processing}. IEEE, 1987, vol.~12, pp. 705--708.

\bibitem{ko2017study}
Tom Ko, Vijayaditya Peddinti, Daniel Povey, Michael~L Seltzer, and Sanjeev
  Khudanpur,
\newblock ``A study on data augmentation of reverberant speech for robust
  speech recognition,''
\newblock in {\em 2017 IEEE International Conference on Acoustics, Speech and
  Signal Processing (ICASSP)}. IEEE, 2017, pp. 5220--5224.

\bibitem{kim2017generation}
Chanwoo Kim, Ananya Misra, Kean Chin, Thad Hughes, Arun Narayanan, Tara
  Sainath, and Michiel Bacchiani,
\newblock ``Generation of large-scale simulated utterances in virtual rooms to
  train deep-neural networks for far-field speech recognition in {Google
  Home},'' 2017.

\bibitem{laptev2020you}
Aleksandr Laptev, Roman Korostik, Aleksey Svischev, Andrei Andrusenko, Ivan
  Medennikov, and Sergey Rybin,
\newblock ``You do not need more data: Improving end-to-end speech recognition
  by text-to-speech data augmentation,''
\newblock {\em arXiv preprint arXiv:2005.07157}, 2020.

\bibitem{gokay2019improving}
Ramazan Gokay and Hulya Yalcin,
\newblock ``Improving low resource {Turkish} speech recognition with data
  augmentation and {TTS},''
\newblock in {\em 2019 16th International Multi-Conference on Systems, Signals
  \& Devices (SSD)}. IEEE, 2019, pp. 357--360.

\bibitem{wang2020improving}
Gary Wang, Andrew Rosenberg, Zhehuai Chen, Yu~Zhang, Bhuvana Ramabhadran,
  Yonghui Wu, and Pedro Moreno,
\newblock ``Improving speech recognition using consistent predictions on
  synthesized speech,''
\newblock in {\em IEEE International Conference on Acoustics, Speech and Signal
  Processing (ICASSP)}. IEEE, 2020, pp. 7029--7033.

\bibitem{oord2016wavenet}
Aaron van~den Oord, Sander Dieleman, Heiga Zen, Karen Simonyan, Oriol Vinyals,
  Alex Graves, Nal Kalchbrenner, Andrew Senior, and Koray Kavukcuoglu,
\newblock ``{WaveNet}: A generative model for raw audio,''
\newblock {\em arXiv preprint arXiv:1609.03499}, 2016.

\bibitem{shen2018natural}
Jonathan Shen, Ruoming Pang, Ron~J Weiss, Mike Schuster, Navdeep Jaitly,
  Zongheng Yang, Zhifeng Chen, Yu~Zhang, Yuxuan Wang, RJ~Skerrv-Ryan, Rif.~A.
  Saurous, Yannis Agiomyrgiannakis, and Yonghui Wu,
\newblock ``Natural {TTS} synthesis by conditioning {WaveNet} on {Mel}
  spectrogram predictions,''
\newblock in {\em 2018 IEEE International Conference on Acoustics, Speech and
  Signal Processing (ICASSP)}. IEEE, 2018, pp. 4779--4783.

\bibitem{ping2018deep}
Wei Ping, Kainan Peng, Andrew Gibiansky, Sercan~O Arik, Ajay Kannan, Sharan
  Narang, Jonathan Raiman, and John Miller,
\newblock ``{Deep Voice} 3: 2000-speaker neural text-to-speech,''
\newblock {\em Proc. ICLR}, pp. 214--217, 2018.

\bibitem{jia2018transfer}
Ye~Jia, Yu~Zhang, Ron Weiss, Quan Wang, Jonathan Shen, Fei Ren, Zhifeng Chen,
  Patrick Nguyen, Ruoming Pang, Ignacio Lopez~Moreno, and Yonghui Wu,
\newblock ``Transfer learning from speaker verification to multispeaker
  text-to-speech synthesis,''
\newblock in {\em Advances in neural information processing systems}, 2018, pp.
  4480--4490.

\bibitem{chen2019sample}
Yutian Chen, Yannis Assael, Brendan Shillingford, David Budden, Scott Reed,
  Heiga Zen, Quan Wang, Luis~C Cobo, Andrew Trask, Ben Laurie, Caglar Gulcehre,
  Aäron van~den Oord, Oriol Vinyals, and Nando de~Freitas,
\newblock ``Sample efficient adaptive text-to-speech,''
\newblock in {\em International Conference on Learning Representations}, 2019.

\bibitem{simonchik2014stc}
Konstantin Simonchik and Vadim Shchemelinin,
\newblock ```{STC Spoofing}' database for text-dependent speaker recognition
  evaluation,''
\newblock in {\em Spoken Language Technologies for Under-Resourced Languages},
  2014.

\bibitem{shchemelinin2013examining}
Vadim Shchemelinin and Konstantin Simonchik,
\newblock ``Examining vulnerability of voice verification systems to spoofing
  attacks by means of a {TTS} system,''
\newblock in {\em International Conference on Speech and Computer}. Springer,
  2013, pp. 132--137.

\bibitem{shchemelinin2014vulnerability}
Vadim Shchemelinin, Mariia Topchina, and Konstantin Simonchik,
\newblock ``Vulnerability of voice verification systems to spoofing attacks by
  {TTS} voices based on automatically labeled telephone speech,''
\newblock in {\em International Conference on Speech and Computer}. Springer,
  2014, pp. 475--481.

\bibitem{kinnunen2017asvspoof}
Tomi Kinnunen, Nicholas Evans, Junichi Yamagishi, Kong~Aik Lee, Md~Sahidullah,
  Massimiliano Todisco, and H{\'e}ctor Delgado,
\newblock ``{ASVspoof} 2017: Automatic speaker verification spoofing and
  countermeasures challenge evaluation plan,''
\newblock {\em Training}, vol. 10, no. 1508, pp. 1508, 2017.

\bibitem{wang2020asvspoof}
Xin Wang and et~al,
\newblock ``{ASVspoof} 2019: {A} large-scale public database of synthesized,
  converted and replayed speech,''
\newblock {\em Computer Speech \& Language}, 2020.

\bibitem{park2019specaugment}
Daniel~S. Park, William Chan, Yu~Zhang, Chung-Cheng Chiu, Barret Zoph, Ekin~D.
  Cubuk, and Quoc~V. Le,
\newblock ``{SpecAugment: A} simple data augmentation method for automatic
  speech recognition,''
\newblock in {\em Interspeech}, 2019.

\bibitem{faisal2019specaugment}
Muhammad~Yusuf Faisal and Suyanto Suyanto,
\newblock ``{SpecAugment} impact on automatic speaker verification system,''
\newblock in {\em International Seminar on Research of Information Technology
  and Intelligent Systems (ISRITI)}, 2019.

\bibitem{wang2020specaugment}
Shuai Wang, Johan Rohdin, Oldřich Plchot, Lukáš Burget, Kai Yu, and Jan
  Černocký,
\newblock ``Application of {SpecAugment} to deep speaker embedding learning,''
\newblock in {\em IEEE International Conference on Acoustics, Speech and Signal
  Processing (ICASSP)}, 2020.

\bibitem{prabhavalkar2015automatic}
Rohit Prabhavalkar, Raziel Alvarez, Carolina Parada, Preetum Nakkiran, and
  Tara~N Sainath,
\newblock ``Automatic gain control and multi-style training for robust
  small-footprint keyword spotting with deep neural networks,''
\newblock in {\em 2015 IEEE International Conference on Acoustics, Speech and
  Signal Processing (ICASSP)}. IEEE, 2015, pp. 4704--4708.

\bibitem{sak2014long}
H.~Sak, A.~Senior, and F.~Beaufays,
\newblock ``Long short-term memory recurrent neural network architectures for
  large scale acoustic modeling,''
\newblock in {\em InterSpeech}, 2014.

\bibitem{kalchbrenner2018efficient}
Nal Kalchbrenner, Erich Elsen, Karen Simonyan, Seb Noury, Norman Casagrande,
  Edward Lockhart, Florian Stimberg, Aaron Oord, Sander Dieleman, and Koray
  Kavukcuoglu,
\newblock ``Efficient neural audio synthesis,''
\newblock in {\em International Conference on Machine Learning}, 2018, pp.
  2410--2419.

\bibitem{hsu2018hierarchical}
Wei-Ning Hsu, Yu~Zhang, Ron~J Weiss, Heiga Zen, Yonghui Wu, Yuxuan Wang, Yuan
  Cao, Ye~Jia, Zhifeng Chen, Jonathan Shen, Patrick Nguyen, and Ruoming Pang,
\newblock ``Hierarchical generative modeling for controllable speech
  synthesis,''
\newblock in {\em International Conference on Learning Representations}, 2018.

\bibitem{kingma2014auto}
Diederik~P Kingma and Max Welling,
\newblock ``Auto-encoding variational bayes,''
\newblock {\em International Conference on Learning Representations}, 2014.

\bibitem{zen2019libritts}
Heiga Zen, Viet Dang, Rob Clark, Yu~Zhang, Ron~J Weiss, Ye~Jia, Zhifeng Chen,
  and Yonghui Wu,
\newblock ``{LibriTTS}: A corpus derived from {LibriSpeech} for
  text-to-speech,''
\newblock {\em Proc. Interspeech 2019}, pp. 1526--1530, 2019.

\bibitem{kim2008robust}
Chanwoo Kim and Richard~M Stern,
\newblock ``Robust signal-to-noise ratio estimation based on waveform amplitude
  distribution analysis,''
\newblock in {\em Ninth Annual Conference of the International Speech
  Communication Association}, 2008.

\bibitem{garofolo_1992_1}
J.~Garofolo, L.~Lamel, W.~Fisher, J.~Fiscus, D.~Pallett, N.~Dahlgren, and
  V.~Zue,
\newblock ``{TIMIT} acoustic-phonetic continuous speech corpus,''
\newblock {\em Linguistic Data Consortium}, 1992.

\end{thebibliography}

\end{document}